\documentclass[onecolumn, amssymb, showpacs]{revtex4-1}				  % use default format for submitting
\usepackage{graphicx}			% Include figure files
\usepackage{dcolumn}			% Align table columns on decimal point
\usepackage{bm}					% bold math
\usepackage{float}			% Include float files         https://www.ctan.org/pkg/float?lang=en
\usepackage{amsmath,amsfonts}	% popular packages from the American Mathematical Society
\usepackage{url}					% https://www.ctan.org/pkg/url
\usepackage{setspace}		% Sets spacing    https://www.ctan.org/pkg/setspace
\usepackage{lineno}   			% get from ctan      https://www.ctan.org/pkg/lineno
\usepackage[colorinlistoftodos]{todonotes}

\begin{document} 

\title{Elastic wave control beyond band-gaps: shaping the flow of waves in plates and half-spaces.}

\author{Andrea Colombi}%
\email[Corresponding author ]{e-mail: a.colombi@imperial.ac.uk}
\author{Richard V. Craster}
\affiliation{Dept. of Mathematics, Imperial College London, South Kensington Campus, London, U.K.}
\author{Daniel Colquitt}
\affiliation{Dept. of Mathematical Sciences, University of Liverpool, Liverpool, U.K.}
\author{Younes Achaoui}%
\affiliation{MN2S, Femto-st Besancon, 25030 Besancon CEDEX France }
\author{Sebastien Guenneau}
\affiliation{Aix-Marseille Universit\'e, Centrale Marseille, Institut Fresnel-CNRS (UMR 7249), 13397 Marseille cedex 20, France}
\author{Philippe Roux}%
\affiliation{ISTerre, CNRS, Univ. Grenoble Alpes, France, BP 53 38041 Grenoble CEDEX 9}
\author{Matthieu Rupin}%
\affiliation{Hap2U, CIME Nanotech, 3 parvis Louis N\'{e}el 38000 Grenoble, France}

\date{\today}

%%%%%%%%%%%%%%%%% END OF PREAMBLE %%%%%%%%%%%%%%%%

\begin{abstract}
It is well known in metamaterials that local resonance and hybridization phenomena dramatically influence the shape of dispersion curves; the metasurface created by a cluster of resonators, subwavelength rods, atop an elastic surface being an exemplar with these features. On this metasurface, band-gaps, slow or fast waves, negative refraction and dynamic anisotropy can all be observed by exploring frequencies and wavenumbers from the Floquet-Bloch problem and by using the Brillouin zone. These extreme characteristics, when appropriately engineered, can be used to design and control the propagation of elastic waves along the metasurface. For the exemplar we consider, two parameters are easily tuned: rod height and cluster periodicity. The height is directly related to the band-gap frequency, and hence to the slow and fast waves,  while the periodicity is related to the appearance of dynamic anisotropy.
Playing with these two parameters generates a gallery of metasurface designs to control the propagation of both flexural waves in plates and surface Rayleigh waves for half-spaces. Scalability with respect to the frequency and wavelength of the governing physical laws allows the application of these concepts in very different fields and over a wide range of lengthscales. 

\end{abstract}

\maketitle
\noindent

\section{Introduction}

%\subsection{general stuff}

Recent years have witnessed the increasing popularity of metamaterial concepts, based on so-called local resonance phenomenon, to control the propagation of electromagnetic \citep{pendry,wiltshire2004,sar12008,werner2016}, acoustic and elastic \citep{Liu,craster2012} waves in artificially engineered media. Initially  attention focused on the existence of subwavelength band-gaps generated by the resonators \cite{pendry1998,movchan2004,cocacola,younes2011,andrea4}, and resulting frequency dependent effective material parameters for negative refraction and focusing effects \cite{Smith2000,pendry2,Yang2002,jensen2004}, and now consideration is transitioning to methods for achieving more complete forms of wave control by encompassing tailored graded designs to obtain spatially varying refraction index \cite{Pendry23062006}, wide band-gaps and mode conversion. In the fields of photonics and acoustics  this transition has already taken place and new graded designs allow for the  tailored control of the propagation of light  \cite{stru_surf,Muamer2011302}, micro-waves \cite{schurig2006}, water waves \cite{PhysRevLett.101.134501} and sound \cite{cummer2007,fang2011,romero13a,sensing}. 
Elastodynamic media have, in contrast to acoustic and electromagnetic systems, additional complexity such as supporting both compressional and shear wave speeds that differ and which mode convert at interfaces \cite{craster2012}. On one hand this makes elastic metamaterials complex to model and require the use of computational elastodynamic techniques \cite{andrea_soil_lenses}, on the other hand it offers new control possibilities not achievable in the electromagnetic or acoustic cases.
Wave control has implications in several disciplines and the discoveries of metasurface science are currently being translated into several applications. 
If we limit our discussion to elastic metamaterials, potential applications can be implemented at any lengthscale. On the large-scale, seismic metamaterials have become very popular \citep{brule,eleni2,finocchio,miniaci_seismic,achaoui_seismic}. At smaller scale, in mechanical engineering, applications based on wave re-routing and protection are currently being explored \citep{andrea_ultrasonic,andrea6} to reduce vibrations in high precision manufacturing and in laboratories for high precision measurements (e.g. interferometry) or in the field of ultrasonic sensing to amplify signal to noise ratio. In the field of acoustic imaging, the tailored control of hypersound (elastic waves at GHz frequencies) used for cell or other nano-compound imaging \cite{della_Picca}, is emerging as one of the most promising applications of energy trapping and signal enhancement through metamaterials. Furthermore, at this small scale, novel nanofabrication techniques deliver the tailoring possibilities required for graded devices \cite[e.g.][]{lucio,redondo}. 

Among the possible resonant metasurface designs for elastic waves proposed in recent years \citep[e.g.][]{Miniaci2015251,Baravelli20136562,Matlack26072016,Lee2016ExtremeSH,Tallarico2017236}, the one made of a cluster of rods (the resonators) on an elastic substrate has revealed superior characteristics and versatility of use in particular towards the fabrication of graded design. 
The physics of this metasurface is well described through a Fano-like resonance \citep{fano}. A single rod attached to an elastic surface couples with the motion of both the $A_0$ mode in a plate or the Rayleigh wave on a thick elastic substrate (half-space). This coupling is particularly strong at the longitudinal resonance frequencies of the rod. At this point, the eigenvalues of the equation describing the motion of the substrate and the rod are perturbed by the resonance and become complex leading to the formation of a band-gap \citep{perkins86a,landau58a}. When the resonators are arranged on a subwavelength cluster (i.e. with $\lambda$, the wavelength $\gg$ than the resonator spacing), as in the metasurface discussed here, the resonance of each rod acts constructively enlarging the band-gap until, approximately, the rod's anti-resonance \citep{matthieu,andrea_tree}. Thus the resulting band-gap is broad and subwavelength. 
Because the resonance frequency of the rod drives the band-gap position, a spatially graded metasurface is simply obtained by varying the length of the rods, which directly underpins the resonance frequency. 
Thus the length of the rod appears to be the key parameter for the metasurface tunability, although the  periodicity and distribution of the rods cannot be ignored as they also influence the dispersion curves leading to zone characterised by dynamic anisotropy and negative refraction \cite{kaina15a}. These effects are important as they may be used to generate highly collimated waves or for subwavelength imaging.    
Our purpose in this work is to complement the research on local resonance and slow and fast waves, with the study of the dynamic anisotropy effect \cite{colquitt2011} when the rods are periodically arranged on the elastic surface.

%\subsection{Additional paragraph on hyperbolic metamaterials from dynamic anisotropy}
In fact, it has been recently realized that many novel features of hyperbolic metamaterials such as superlensing and enhanced spontaneous emission \cite{poddubny2013} could be achieved thanks to dynamic anisotropy in photonic \cite{ceresoli2016} and phononic crystals \cite{colquitt2011,mouldin_waves}. For instance, the high-frequency homogenization theory \cite{Craster01112010} establishes a correspondence between anomalous features of dispersion curves on band diagrams with effective tensors in governing wave equations: flat band and inflexion (or saddle) points lead to extremely anisotropic and indefinite effective tensors, respectively, that change the nature of the wave equations (elliptic partial differential equations can turn parabolic or hyperbolic depending upon effective tensors). This makes analysis of dynamic anisotropy a potentially impactful work.

The first half of the article is dedicated to the review of the state of the art on the control of flexural and Rayleigh waves with rods on an elastic substrate. This part will collect the major achievements and milestones obtained by our research group in the past 3 years. In the second part we will present another characteristic of this metasurface analysing the 2D dispersion curves and the effect of dynamic anisotropy in the subwavelength regime.

\section{Early results: plate vs. infinite half-space metamaterial}
We start by recalling results obtained with a metamaterial, introduced in 2014, made from a thin elastic plate and a cluster of closely spaced resonators (see model in Fig.~1(a)) both made of aluminium. At that time, despite the limited knowledge of the metasurface dispersion properties, the cluster of resonators immediately showed surprising phenomena such as the presence, in the Fourier spectra, of large subwavelength band-gaps \citep{matthieu} affecting the propagation of the $A_0$ mode in the thin plate in the kHz range. 
Around the same time, \cite{andrea4} demonstrated that by exploiting the stop band,  waves can be trapped in a very subwavelength cavity and that energy could be tunneled through the metasurface by inserting a defect, with an approach reminiscent of phononic crystal applications. 
These early attempts to compute the dispersion curves of the metasurface for a given rod size and spacing were based on array methods that projected the time series recorded from either experiment or numerical simulation on the frequency wavenumber plane ($f-k$ plane). These preliminary results confirmed the resonant nature of the band-gap and uncovered another striking characteristic of the metamaterial: the nearly flat branches occurring at edges of the Brillouin zone before and after the band-gap. These flat branches represent, for high wavenumbers, very slow modes. Conversely for $k$ approaching the origin, these modes travel very fast. 
The analytical calculation of the dispersion properties by \cite{williams} (e.g. the plot in Fig. 1(a)), mean we can now fully harness the power of this metasurface and use the concept of fast and slow modes to fully control the propagation of waves in a plate \citep{andrea6}. Before showing the effects of such tailored wave control we continue our digression into the important applications of elastic resonators on an elastic surface. It has been known since \cite{khelif_pillars1} that short pillars (or other type of resonators \citep{PhysRevLett.111.036103}) on an elastic half-space can  alter the dispersion curves by introducing Bragg and resonant band-gaps for Rayleigh waves. However, the use of longitudinally elongated resonators, such as the rods shown in Fig 1b, allow for a much clearer  separation of the longitudinal mode (responsible for the band-gap) from other flexural resonances that will be discussed in the last section of this article. This has the twofold advantage of pushing the band-gap to the subwavelength scale, simultaneously increasing its breadth, and also simplifying the analytical description of the metamaterial. 
From an analytical point of view the thin elastic plate metasurface can still be treated as a scalar problem as one can use Kirchhoff's plate theory coupled with a longitudinal wave equation for the rod. In an elastic half-space this is no longer possible and the full elastic equation must be used to describe its physics. With this concept in mind \citep{daniel_andrea} constructed an analytical formulation for the dispersion curve of a 1D array of resonators on the half-space considering only the longitudinal modes of the rods. 
From visual inspection of the plot in Fig. 1(b), besides the obvious lack of dispersion for body and Rayleigh waves in the half-space (in contrast the $A_0$ mode in Fig. 1(a) is strongly dispersive) and the different frequency and size of the model (metres and kHz for the plate and centimetres and MHz for the half-space), a similar hybridization mechanism \citep{fano} creates the band-gap in both systems. However, in the half-space, the maximum speed of the system is bounded by the shear S-wave line. These observations are consistent with the physical interpretation that the vertical component of the elliptically polarised Rayleigh waves, usually travelling  slower then the shear wave, couples with the longitudinal motion of the resonator.
The presence of these band-gaps have inspired the development of so-called seismic metamaterials for Rayleigh waves \cite{andrea_tree} where the close relationship between shear S and Rayleigh waves in the half-space lead to unexpected wave phenomena in the metamaterial. As chiefly demonstrated in \citep{colombi16a} and \cite{daniel_andrea}, the resonance creates an hybrid branch bridging the Rayleigh line with the S-wave line. Through a graded resonators design (e.g. decreasing or increasing rod's height) the conversion becomes ultra-broadband, a key requirement for practical engineering applications. 

\section{Gallery of control possibilities achieved by tuning the rod length}
The rich physics encoded within the hybrid dispersion curves that we have just described for the plate and half-space cases, can be translated into extraordinary wave propagation phenomena. Furthermore, scalability is one of the strong characteristics of metamaterials which makes them applicable in different wave realms and lengthscales. With the following examples we  demonstrate that applications for the two different settings and lengthscale introduced in Figs. 1(a) and (b), namely the elastic plate and the half-space. This choice is made to remain coherent with our previous laboratory and numerical studies on these structures \cite{matthieu,andrea_ultrasonic}.
The description starts from Fig. 2(a), snapshots extracted from a numerical simulation show the band-gap created by a small cluster of resonators located on top of a thin elastic plate. The field has been filtered inside the band-gap at a frequency between 2 and 3 kHz (6 mm thick plate and 60 cm long rods, both made of aluminium). The band-gap frequency $f$ directly depends on the resonator length $h$ and therefore can be easily tuned by selecting longer or shorter rods using the well known formula:
\begin{equation}
\label{eqn:risonanza}
f=\frac{1}{4 h}\sqrt{\frac{E}{\rho}},
\end{equation} 
where  $E$ its Young's modulus and $\rho$ its density. This formula is valid when the substrate is sufficiently stiff, for seismic metamaterials, where the resonator might be supported by a soft sediment layer, the contribution of the substrate must be taken into account when calculating the resonance frequency \citep{andrea_tree}. 
 
In Fig. 2(b) and (c), we exploit the effective wave velocity (slow waves) that is locally achieved in the metamaterial. In these figures we show two types of so-called graded index lenses \citep{2040-8986-14-7-075705} well known for their capacity to focus and re-route waves without aberration and reflection. These lenses are characterised by a radially varying velocity profile decreasing from the outside to the inside (for 4 kHz flexural waves $h$ varies approximately between 60 and 80 cm while  in Fig. 2(b) while between 60 and 90 cm for the case in Fig. 2(c)). In practice, such a material is very difficult to fabricate unless one uses layers of different material or a graded thickness profile for the plate case. For the half-space this is clearly not possible. By using the slow modes of the flat branch occurring before the band-gap (see dispersion curves in Figs. 1(a) and (b)) these velocity gradients can be achieved by tailoring the resonator height distribution to the velocity profile required by the lens. This step is better achieved using the analytical form of the dispersion curve as shown in \cite{andrea6} derived using the theory from \cite{williams}.
Although only the results for the plate have been currently published, the same method can be applied to Rayleigh waves too with the theory developed by \cite{daniel_andrea}. 

In the remaining three figures the description moves to the control of Rayleigh waves. Unlike the plate case where the physics can be captured in the plane, here it is important to describe the whole  wavefield inside the half-space. For this reason 2D simulations in the $P-SV$ plane (plane  strain) are now shown. 
As already anticipated in Fig. 1(b), the first snapshot shows the band-gap (here the field is filtered between 0.35-0.4 MHz) produced by an array of resonators of constant height. In Fig. 2(e) we show the well-known phenomena of the rainbow trapping but now for elastic waves. As for the lens case, this effect is completely due to the slow branch occurring below the band-gap. The graded array of resonator (resonant metawedge), enhances this effect and makes this device completely broadband (inversely proportional to the height). Note that,  compared to the band-gap described in Fig. 2(d), here the Rayleigh wave remains confined to the surface while a broadband band-gap is produced after the wedge; for clarity of presentation we have used a monochromatic source of Rayleigh waves at 0.5 MHz.
When the wedge orientation is reversed, as in Fig. 2(f), the surprising phenomenon of modal conversion is obtained and the graded profile enhances the conversion on a large frequency band; in the previous section this was already anticipated from the analysis of the dispersion curves. 
The control possibilities emerging from this discussion suggest tremendous potential for applications of these metamaterials toward vibration reduction and enhanced sensing. 
%In civil engineering: from ground borne vibration reduction to the mitigation of the effect of surface waves induced by earthquake. At smaller scale we can envisage their usage in high precision manufacturing, sensing or, as already discussed, in the tailored control of hypersounds for acoustic imaging. 
In this section we have not specified yet whether these phenomena depend, or not, on the periodicity of the resonator distribution in the metamaterial. 
Because local resonance is at the origin of the effects presented so far the answer is no for all of them. However, in the next section we will explore the important implications of periodicity. 

\section{Periodicity, dynamic anisotropy and hyperbolic behaviour}
The height of the rods is not the only parameter available in terms of design of the metasurface. Solid state physics informs us that the lattice periodicity and spacing also matter as that generates, in particular, Bragg-type scattering. Dynamic anisotropy, that is anisotropy observed in the wavefield, that changes as frequency varies, is a common feature in phononic crystals with the most extreme situation being that where the wave energy is confined to ``rays'' with the field taking a cross-like form. Despite this, it has only marginally been associated with subwavelength metamaterials \cite{kaina15a,maznev15a} with most work carried out in the context of phononic crystals. This section explores how anisotropy is obtained with this metasurface design. We introduce in Figs. 3(a) and (b) the dispersion curves for a 2D array of resonators respectively on a plate and on an infinite elastic support (half-space). The analysis is carried out inside the well-known irreducible Brillouin zone defined on the wavevector plane $\mathbf{k}$=($k_x$,$k_y$) by the three points of coordinats: $\Gamma=(0,0)$, $M=(\pi/d,\pi/d)$ and $X=(\pi/d,0)$ where $d$ is the pitch of the array of resonators. Given the complexity of the 3D problem, the model is solved numerically and includes all the admissible modes of the unit cell, not only, as previously done, the elongation of the rods. The resulting dispersion curves are characterised by several resonances that makes it hard to distinguish the longitudinal one. To aid interpretation we plot, along with the curve, the ratio between the vertical and the longitudinal value of the eigenfunction measured at the top of the resonator (where for all modes, the displacement reach a maximum \citep{modal_test}). High values mean that the motion is vertically polarised, conversely low value means that motion is horizontal; this interpretation is further confirmed associating to each resonance its modal deformation. 

The size of the unit cell in Fig. 3(a) is chosen to be similar to the cluster configuration in our previous work \cite{andrea4,matthieu} where we have used a 6~mm plate and 60~cm rods both made of aluminium. The eigenvalue analysis is done using COMSOL and we make use of the built-in Bloch-Floquet boundary conditions to mimic an infinite 2D array of rods that are  3-cm-spaced.  The bare plate dispersion curve is shown in red for the $\Gamma$-X direction that is equal to the configuration in Fig. 1(a) (although without flexural resonances). Thanks to the colorcode used, the longitudinal modes are clearly identified in the dispersion curves. Given the lattice size, the first longitudinal mode is very subwavelength $\sim\lambda/8$. While the zoomed detail around this resonance is shown in Fig. 4, we can already distinguish the change in curvature that is responsible for the dynamic anisotropy behaviour. The other flat branches are mainly flexural modes (except for some breathing mode of the resonator). These are all double modes because the resonator is free to move in both directions.
In Fig. 3(b) we repeat the same analysis for the half-space. The dimensions of the unit cell are similar to those for the plate although the spacing is slightly larger to improve the visualization of the anisotropy in Fig. 4(b). A technical detail is that, 
to mimic the infinite character of the half-space, we have applied an absorbing boundary at the lower side of the computational cell (see COMSOL Structural Mechanics Module documentation). 
The physics of the wave propagation in the half-space differs from the plate case mainly because of the lack of dispersion (see the straight dispersion curves for the bare half-space) and the higher speed of the waves. In this configuration however, the longitudinal resonance is only slightly subwavelength $\sim\lambda/3$. Clearly, by using a longer resonator, the band-gap can be pushed to a much lower frequency but the curvature of the longitudinal resonance is then shrunk down to a fraction of the Hertz, making the visualization  of the anisotropy practically impossible as the effect is so sensitive that small numerical or manufacturing variations would spoil the expected result. 

We now focus on the anisotropic behaviour by zooming in to frequencies close to the longitudinal modes. A detailed view of the first mode of the plate is depicted in Fig. 4(a). We can clearly appreciate the slope change that occurs before the resonance. A spectral element simulation in the time domain shows a snapshot of the wavefield filtered around the inflexion point of the mode. An array of $20\times20$ resonators, spaced and sized according to Fig. 3(a), is placed at the center of the plate. Because the plate boundaries are reflecting, to improve the visualization despite the reverberations, we have smoothed the square array removing the corner. The cluster is in fact octagonal. The shape and size of the plate is identical to the one used in \cite{matthieu}, so this phenomenon could be easily verified experimentally. The source is located in the middle of the array and in our case it is broadband Gaussian pulse. 
The cross-shaped anisotropic profile, as well as the strong contrast between the wavelength inside and outside the plate is clearly visible, and reminiscent of wave patterns in negatively refracting and hyperbolic metamaterials. 

Using the same modeling technique, dynamic anisotropy also characterizes the half-space and it is indeed visible in the numerical results of Fig. 4(b). The half-space is simulated applying perfectly matched layers on the side and on the bottom surface. The snapshot show the vertical component of the displacement filtered at the inflexion point. As for the case of the plate, a similar cross is visible. However, here we observe a strong spatial attenuation of the field due to the fact that waves are free to propagate or scatter downward, while in the plate they were guided (e.g. Fig. 2(d)).  

At this stage, we note that there is a vast literature on electromagnetic hyperbolic metamaterials, which were theorized by David Smith and David Schurig almost fifteen years ago in the context of negatively refracting media described by electric permittivity and magnetic permeability tensors with eigenvalues of opposite signs \cite{smith2003,smith2004}. These media originally thought of as an anisotropic extension of John Pendry's perfect lens \cite{pendry2,luo2002} take their name from the topology of the isofrequency surface. In an isotropic homogeneous medium (e.g. vacuum in electromagnetics and air in acoustics), the linear dispersion and isotropic behaviour of transversely propagating (electromagnetic or sound) waves implies a circular isofrequency contour given by the dispersion equation $k_x^2+k_y^2=\omega^2/c^2$ with ${\bf k}=(k_x,k_y)$ the wavevector, $\omega$ the angular wave frequency and $c$ the wavespeed of light or sound waves. In a transversely anisotropic  effective medium, one has $T_{yy} k_x^2+T_{xx} k_y^2=\omega^2/c^2$, where $T_{xx}$ and $T_{yy}$ are entries of the (inverse of) effective tensor of permittivity or mass density, shear or Young moduli etc. depending upon the wave equation. It is well known that the  circular isofrequency contour of vacuum distorts to an ellipse for the anisotropic case. However, when we have extreme anisotropy such that $T_{xx}T_{yy}<0$ the isofrequency contour opens into an open hyperbole. In electromagnetics, such a phenomenon requires the metamaterial to behave like a metal in one direction (along which waves are evanescent) and a dielectric in the other and similarly, in acoustics and platonics. A hallmark of hyperbolic media is an X-shape wave pattern for emission of a source located therein \cite{poddubny2013}, reminiscent of the hyperboles arising from the dispersion relations. Note of course, that if both entries of the effective tensor are negative, this means waves are evanescent in all directions, what corresponds to a metal in electromagnetics. 
% we use it for the hyperbolic material paper this! A bit too technical here.
%Importantly, unlike for usual media described by positive definite tensors, hyperbolic media have wavevector solutions that can have an arbitrarily large norm (think of the branches of the hyperboles that go to infinity in the dispersion equation). This means that one can achieve control of light/sound at very small wavelengths (e.g. for superlensing \cite{lu2012}). Moreover, the density of states diverges in such media, and this leads to applications in efficient aborbers as well as controlled spontaneous emission \cite{noginov2010} and enhanced Purcell effect \cite{poddubny2013}, and it is noticeable that a metamaterial originally designed for directive emission \cite{enoch2002} is now used for enhancement of photon density of states.

In the case of structured plates, as aforementioned the high-frequency homogenization theory predicts that at the inflexion point  in Fig. 4(a), the effective tensor (that encompasses an anisotropic Young's modulus) has eigenvalues of opposite sign in the framework of the simplified model of Kirchhoff-Love \cite{mouldin_waves}, and thus Lamb waves propagate like in a hyperbolic medium, and the similar features observed in Fig. 4(b) lead to an analogous conclusion for Rayleigh waves in structured halfspaces. 

%Same as before, we will do this in the hyperbolic paper 
%This means that if we now place the source outside the array of rods, and provided that we manage to avoid reflection induced by impedance mismatch, we have all the ingredients to design a flat convergent lens akin to Pendry's perfect lens, but here for Rayleigh waves. Another interesting potential application suggested by analogies with photonics is that of extraordinary transmission, which can be seen as a form of cloaking, if one sends a plane wave on this array of rods at frequency close to the inflexion point. 

\section{Future perspectives}

Devices based upon exploiting band-gap phenomena, as seismic shields using ideas from Bragg-scattering \cite{brule,miniaci_seismic} or zero-frequency stop-bands \cite{achaoui_seismic}, are gaining in popularity. Given the nuisance of ground vibration, and the importance of elastic wave control, for the urban environment this will be an area of growing importance; the additional degrees of freedom, control over sub-wavelength behaviour, and the broadband features that can be utilised using the resonant sub-wavelength structures discussed herein make them very attractive alternatives. At smaller scale one moves toward the manipulation of mechanical waves in vibrating structures, again it is the long-wave and low frequency waves that one often wants to control and, again, these are precisely the waves that are targeted by sub-wavelength resonator array devices. The ability to spatially segregate waves by frequency, the field enhancement and potential to mode convert surface to bulk waves, Fig. \ref{fig:2}(d-f), are all phenomena with practical importance. Similarly, the ability to control surface waves to create concentrators, surface lenses and to redirect waves, using sub-wavelength arrays, Fig. \ref{fig:2}(a-c), are powerful examples to draw upon for devices. 
The combined features of a flat band and a change of curvature near the inflexion point in Fig. 4 means
that we are in a position to achieve effective parameters with eigenvalues of opposite sign exhibiting very different absolute values. So one can imagine controlling Rayleigh waves that would undergo simultaneously positive and negative refraction on the subwavelength scale, and this could lead to cloaking devices analogous to hyperbolic cloaks in electromagnetics \cite{kim2015}. At the geophysics scale applications of hyperbolic cloaks for Rayleigh waves are in seismic protection. It has been also suggested that one can achieve black hole effects \citep{krylov_black_holes} in hyperbolic metamaterials \cite{Smolyaninov2012}, and this would have interesting applications in energy harvesting for Rayleigh waves propagating through arrays of rods at critical frequencies.

Given the relative youth of metamaterials, as a field, and the very recent translation of metamaterial concepts to elastic plate, and elastic bulk, media there are undoubtedly many phenomena that will translate across from the more mature optical metamaterial field. Metasurfaces have become popular in optics as they can be created to combine the vision of sub-wavelength wave manipulation, with the design, fabrication and size advantages associated with surface excitation. These powerful concepts, and the degree of control available, is driving progress in optics towards flat optical lenses and devices \cite{yu14a}; the elastic analogues of these optical metasurfaces are those we describe here and we anticipate similar progress in the design of mechanical devices.

\bibliographystyle{jasa-ml}
%\bibliography{scibib}%,biblio_meta}

\section*{Acknowledgements}
All of the computations presented in this paper were performed using the Froggy platform of the CIMENT infrastructure (\textit{https://ciment.ujf-grenoble.fr}), supported by the Rhone-Alpes region (GRANT CPER07$\_$13 CIRA), the OSUG2020 labex (reference ANR10 LABX56) and the EquipMeso project (reference ANR-10-EQPX-29-01) of the programme Investissements d'Avenir supervised by the Agence Nationale pour la Recherche.
A.C and R.C. thanks the EPSRC for their support through research grant
 EP/L024926/1. A.C. was supported by the Marie Curie Fellowship ''Metacloak". A.C., P.R., S.G and R.C. acknowledge the  support of the French project Metaforet (reference ANR) that facilitates the collaboration between Imperial College, ISTerre and Institut Fresnel.

\begin{figure}
\centering{}\includegraphics[clip,width=18cm,trim = 0mm 0mm 0mm 0mm]{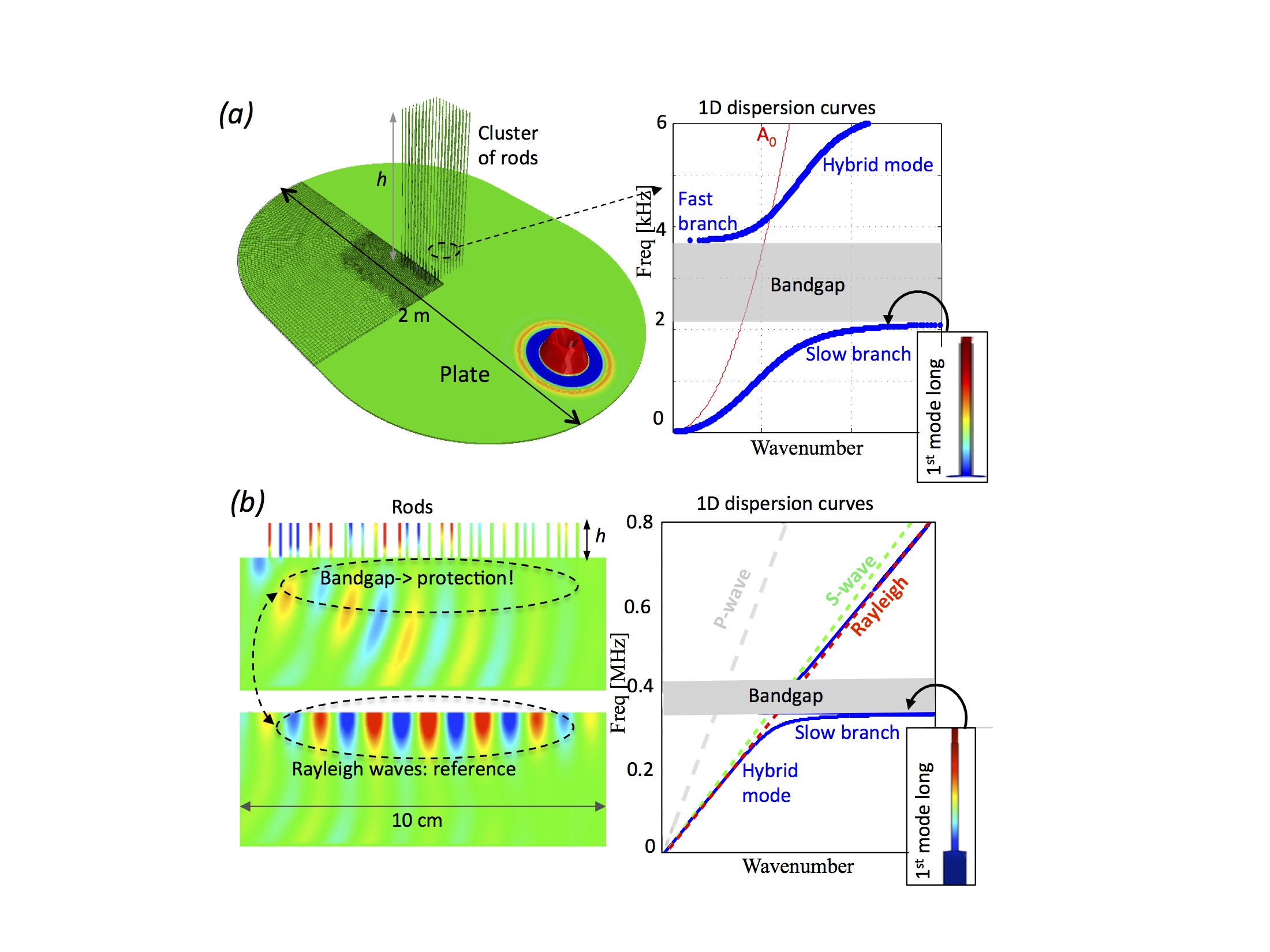}
\protect\caption{(a) The metamaterial discussed in this article is made of a plate and a cluster of closely spaced rods in the same material. It is characterized by a hybrid dispersion curve (blue line), very different from the $A_0$ mode (red line) that propagates in a reference pristine plate excited by a vertical force. 
The plate and the rods are made of aluminum. The inset shows the fundamental longitudinal mode shape that generates the hybrid curve and the band-gap (shaded region). In this formulation, flexural modes are neglected and the colorcode represents the displacement. (b) Same as (a) but for second type of metamaterial discussed here: A 2D elastic half-space where rods are attached to the top surface creating a band-gap for Rayleigh waves. Contrarily to the plate case, the reference half-space is characterized by non-dispersive Rayleigh (red), S-wave (green) and P-wave (gray) dispersion curves. 
 \label{fig:1}}
\end{figure}

\begin{figure}
\centering{}\includegraphics[clip,width=18cm,trim = 0mm 0mm 0mm 0mm]{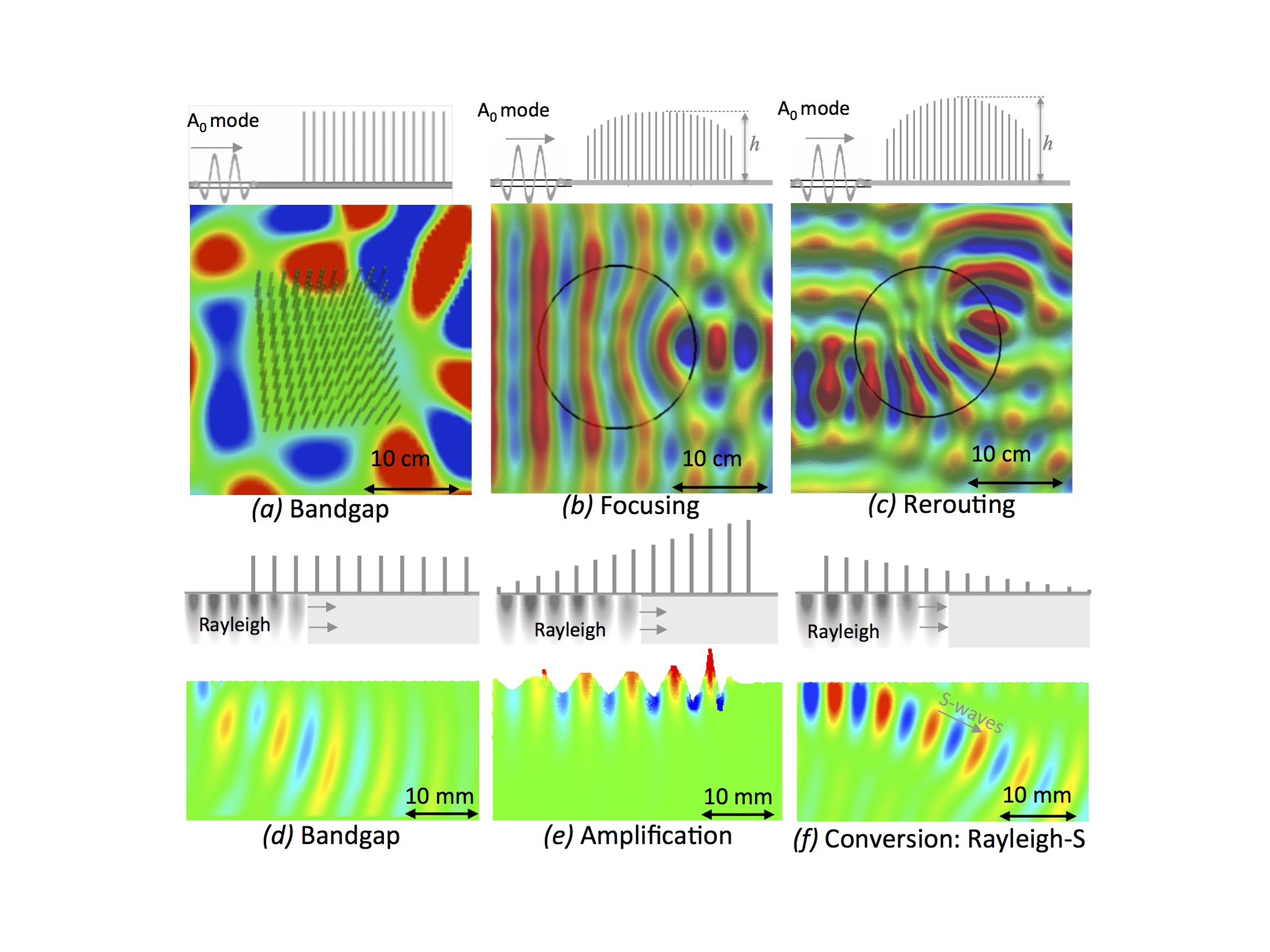}
\protect\caption{ Examples of control achieved through this resonant metamaterial for different types of waves (Flexural or Rayleigh) and lengthscale.
 Colorcode represents the displacement. (a) Band-gaps can stop the propagation of flexural and Rayleigh waves leaving desired regions free of vibrations. (b-c) Elastic energy can be guided or focused with gradient index lenses. (d) Rayleigh wave band-gap created by the constant height resonators. (e) Waves can be spatially segregated depending on the frequency and strongly amplified. (f) Rayleigh waves can be converted to S-waves and redirected in the bulk.  The aspect ratio of the rods and the height gradient are not not in scale to better present the concept. \label{fig:2}}
\end{figure}
\newpage
\begin{figure}
\centering{}\includegraphics[clip,width=18cm,trim = 0mm 0mm 0mm 0mm]{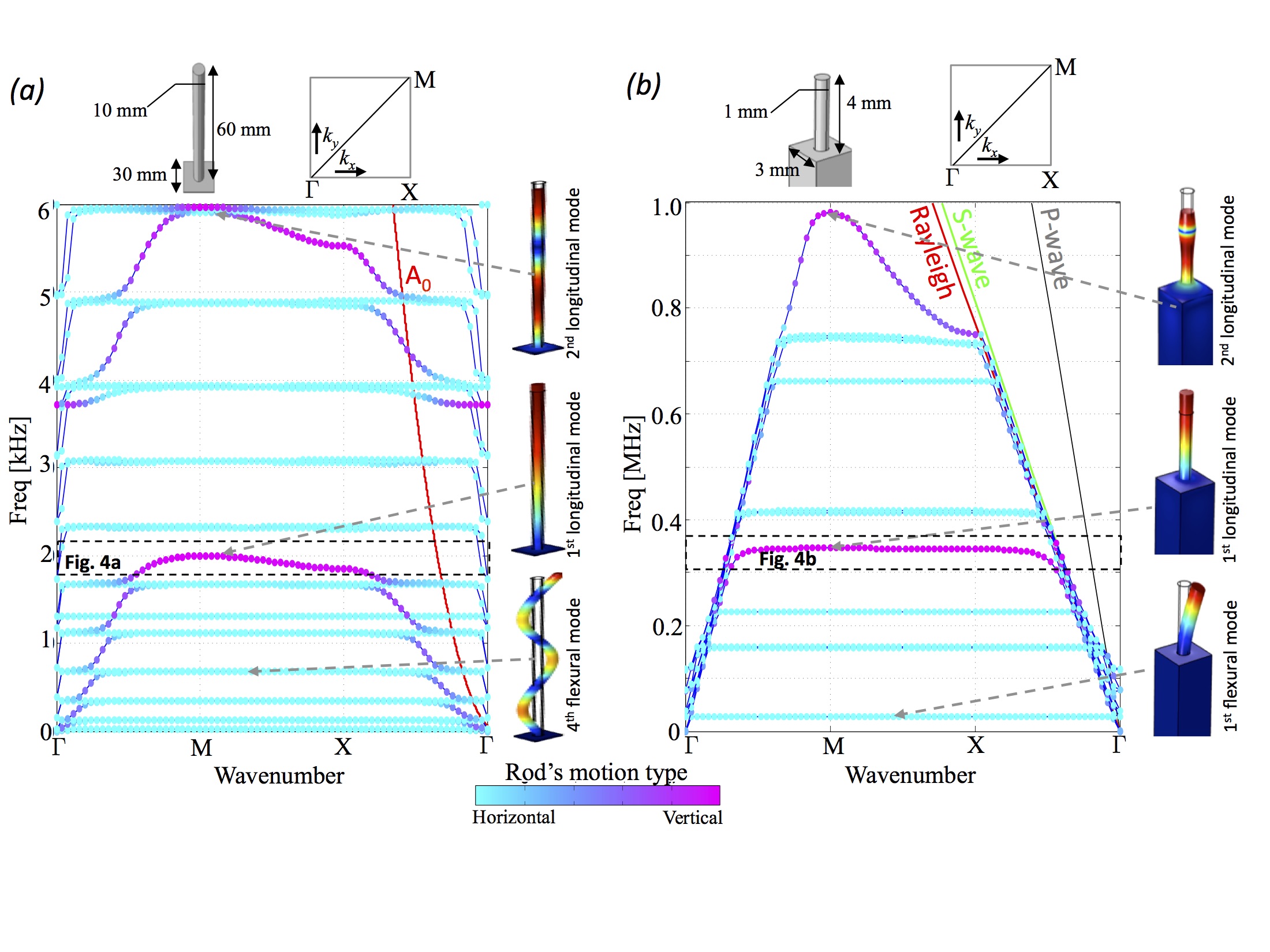}
\protect\caption{(a) Complete dispersion curves for an infinitely periodic 2D layout of rods on an elastic plate.  The size of the unit-cell and the irreducible Brillouin zone ($\Gamma-M-X-\Gamma$) is given at the top. The colorcode superimposed on the dispersion curves represent the motion polarization of the rod. The bare-plate reference $A_0$, has been relocated in the same crystallographic direction as Fig 1a. Snapshots from frequency domain numerical simulation clarify the modal deformation associated to each resonance.  (b) Same as (a) but for the half-space. Here, the propagative zone is bounded by the S-wave (green) maximum velocity (analogous of the so-called light-line in plasmonics).
 \label{fig:3}}
\end{figure}
\newpage
\begin{figure}
\centering{}\includegraphics[clip,width=18cm,trim = 0mm 0mm 0mm 0mm]{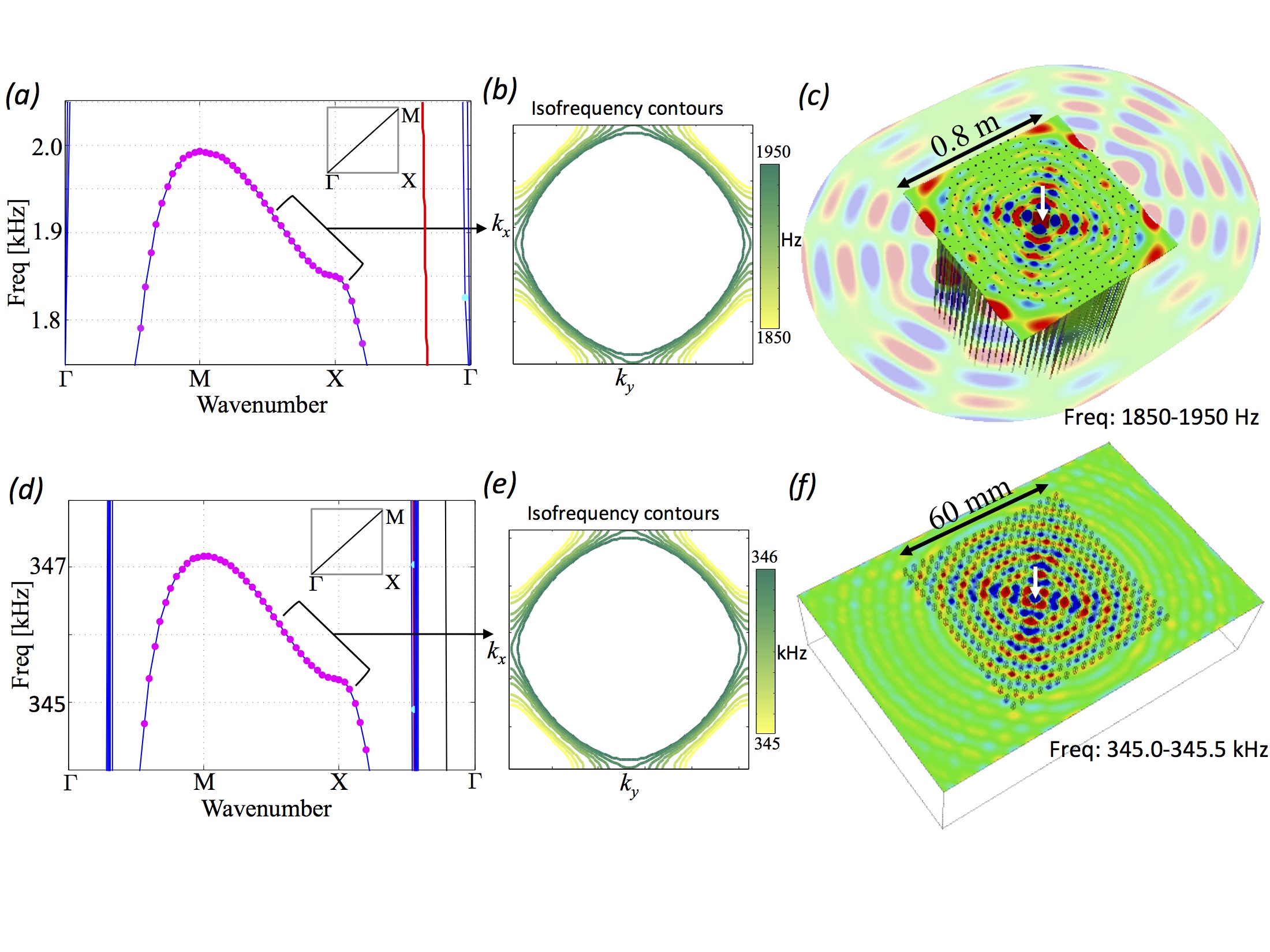}
\protect\caption{(a) A zoomed section around the first longitudinal resonance of the dispersion curves in Fig. 3(a) reveals the strong dynamic anisotropy of this region. (b) The isofrequency contours in the $k_y=(-\pi/d,\pi/d)$ and $k_y=(-\pi/d,\pi/d)$ space show the hyperbolic behaviour of the system around the inflexion point.
(c) Snapshot taken from a time domain numerical simulation with the source located at the center of a cluster of rods. The field has been filtered in the band point by the arrow in the dispersion curve plot. The colorscale represents the vertical component of the displacement field. To ease the visualization of the anisotropic pattern, the plate is represented from the backside and a transparency filter is applied outside the metamaterial to show both the rods and the field in the bare plate. (d-f) Same as (a-c) but for the rods cluster on a half-space. In this case, we only show the vertical displacement field on the top surface. while the actual shape and dimension of the numerical model is sketched with thin black lines. 
 \label{fig:4}}
\end{figure}

\end{document}